\documentclass{IEEEtran}

\usepackage[nocompress]{cite}
\usepackage{verbatim}
\usepackage{amssymb}
\usepackage[cmex10]{amsmath}
\usepackage[amsmath,amsthm,thmmarks]{ntheorem}
\usepackage{algorithmic}
\usepackage{graphicx}
\usepackage[tight,normalsize,sf,SF]{subfigure}
\usepackage{color}
\usepackage{url}
\usepackage{varioref}

\newtheorem{definition}{Definition}
\newtheorem{proposition}{Proposition}

\usepackage{tikz}
\usetikzlibrary{arrows,shapes,snakes}
\usepackage{pgfplots}

\usepackage{sidecap}
\usepackage[normalem]{ulem}

\begin{document}

\title{Adaptive Redundancy Management for Durable P2P Backup}

\author{Matteo~Dell'Amico,
  Pietro~Michiardi,
  Laszlo~Toka,
  and Pasquale~Cataldi
}

\maketitle

\begin{abstract}
We design and analyze the performance of a redundancy management
mechanism for Peer-to-Peer backup applications. Armed with the
realization that a backup system has peculiar requirements -- namely,
data is read over the network only during restore processes caused by
data loss -- redundancy management targets \emph{data durability}
rather than attempting to make each piece of information availabile at
any time.

In our approach each peer determines, in an on-line manner, an amount
of redundancy sufficient to counter the effects of peer deaths, while
preserving acceptable data restore times. Our experiments, based on
trace-driven simulations, indicate that our mechanism can reduce the
redundancy by a factor between two and three with respect to
redundancy policies aiming for data availability. These results imply
an according increase in storage capacity and decrease in time to
complete backups, at the expense of longer times required to restore
data. We believe this is a very reasonable price to pay, given the
nature of the application.

We complete our work with a discussion on practical issues, and their
solutions, related to which encoding technique is more suitable to
support our scheme.

\end{abstract}

\section{Introduction}
\label{sec:introduction}

Online storage solutions are an extremely successful
way of sharing and syncronizing data between machines, taking
advantage of the ubiquity of Internet connectivity. Dropbox, Google
Drive and Microsoft SkyDrive are only a few widely used examples
within the plethora of applications that give this kind of service.

The above mentioned applications adopt a centralized ``cloud''
architecture, with all data residing on the data centers of a single
vendor.  Despite its success, such an architecture has some intrinsic
shortcomings. Some of them have already shown up in news: data loss
due to correlated failures~\cite{carbonite}, security blunders due to
configuration errors~\cite{dropbox}. Others, such as data theft from
rogue employees, might happen eventually. Also, we argue
that \emph{long-term} storage is a case where the weaknesses of
centralized storage are most important: indeed, the costs of storing
large amounts of data over long periods are high, and services might
shut down in the future as already happened to Drop.io~\cite{dropio},
making data availability in the long run essentially impossible to
evaluate.

Peer-to-peer (P2P) storage could solve these problems, providing cheap
storage leveraging on excess bandwidth and disk space at the edge of
the network. However, despite a considerable amount of research, P2P
storage solutions failed to reach widespread usage. Indeed,
implementing a generic P2P storage application requires dealing with a
variety of challenging problems, such as scalable handling of
metadata, dealing efficiently with maintenance due to disk crashes,
low-latency access and modification to individual files, and security
issues such as ensuring data confidentiality even when usage
permissions change dynamically. In this setting, maintaining data
availability in a situation of high churn is a daunting
task~\cite{blake-rodrigues03}.

We take a pragmatic approach: rather than trying to solve all the
aforementioned issues at once and come up with a generic P2P
file-system for the Internet, we design a system exclusively
for \textit{data backup}. Indeed, we argue that backup is a widely
needed application that better fits the characteristics provided by
P2P architectures. Costs and poor usability are among the main reasons
why many existing backup solutions are not used: a P2P approach to
data backup can be a viable technique to overcome such issues.

For backup applications, as we discuss in
Section~\ref{sec:motivations}, the focus shifts from data availability
to durability, which amounts to guaranteeing that data is not
lost. The requirements for a specialized backup application are less
stringent than those of generic storage in several aspects. First,
backups should only be readable by their owner; this makes
confidentiality requirements easy to satisfy with standard
cryptographic techniques. Second, data backup often involves the bulk
transfer of potentially large quantities of data, both during regular
backups and, in the event of data loss, during restore operations.
Therefore, read and write latencies of hours have to be tolerated by
users. Third, owners have access to the original copy of their data,
making it easy to inject additional redundancy in case data stored
remotely is partially lost. Fourth, since data is read only during
restore operations, the application does not need to guarantee that
any piece of the original data should be promptly accessible in any
moment, as long as the time needed to restore the whole backup remains
under control.

In this work, we design and evaluate a new \textit{redundancy
  management} mechanism tailored to backup applications. Simply
stated, the problem of redundancy management amounts to computing the
necessary redundancy level to be applied to backup data to achieve
durability. The endeavor of this work is to design a mechanism that
achieves data durability without requiring high redundancy levels nor
fast mechanisms to detect node failures. Our solution to the problem
stems from the particular data access workload of backup applications:
data is written once and read rarely. The gist of our redundancy
management mechanism, which is described in Section~\ref{sec:redundancy},
is that the redundancy level applied to backup data is computed in an
\textit{on-line} manner. Given a time window, that accounts for
failure detection and data repair delays, and a system-wide statistic
on peer deaths, a peer determines the redundancy rate during the
backup phase. A byproduct of our approach is that, if the system state
changes, then peers can adapt to such dynamics and modify the
redundancy level on the fly.

The ability to compute the redundancy level in an \textit{on-line}
manner requires solving several problems related to coding efficiency
and data management. In Section~\ref{sec:coding}, we show how our scheme
can be realized in practice, exploiting the properties of Fountain
Coding.

Finally, we evaluate our redundancy management scheme using
trace-driven simulations. In Section~\ref{sec:evaluation}, we show that
our approach drastically decreases strain on resources, reducing the
storage and bandwidth requirements by a factor between two and three,
as compared to redundancy schemes that use a fixed, system-wide
redundancy factor. This result yields augmented storage capacity for
the system and decreased backup times, at the expense of increased
restore times, which is a reasonable price to pay if the specific
requirements of backup applications are taken into account.

\section{Application Scenario}
\label{sec:motivations}
        Similarly to many online backup applications, we assume users
(referred to as \emph{data owners}) to specify one local folder
containing important data to backup. Note that backup data remains
available \textit{locally} to data owners. This is an important trait
that distinguishes backup from many online storage applications, in
which data is only stored remotely.

We consider here the problem of long-term storage of large, immutable,
and opaque pieces of data that we term \emph{backup objects}. They
consist of encrypted archives of changes to sets of files, such that
recovering them allows reconstructing the history of data in the
backup folder. We do not take into account the short-term storage of
small modifications to the backup folder, which can be handled using
known centralized or decentralized online storage solutions.

Backup objects are stored on remote peers, which are inherently
unreliable. Peers may join and leave the system at any time, as part
of their short-term online behavior: in the literature, this is referred
to as \textit{churn}. Moreover, peers may crash and possibly abandon
the P2P application: this behavior is generally referred to as peer
\textit{death}. As such, the online behavior of peers must be
continuously tracked, since it cannot be determined \textit{a priori}
\cite{Kiran04totalrecall}.

While the literature provides a vast array of solutions to guarantee
\textit{data availability} when using failure-prone machines to store
data \cite{Kiran04totalrecall, farsite}, we claim that online data
backup applications should instead target \textit{data durability}.
Moreover, backup applications often involve the bulk transfer of a large
quantity of data. Therefore, such applications should cater throughput
rather than aiming at low-latency read operations, in addition to be
resilient against peer churn and deaths. 

Similarly to data availability, data durability can be achieved by
injecting a sufficient level of redundancy in the system. One key
issue to address is to determine the redundancy level required to make
sure data is not lost, despite peer churn. This problem is called
\textit{redundancy management}. A closely related problem is to deal
with peer deaths, which cause the data redundancy level to drop.
Hence, the focus of our work is to design a redundancy management
mechanism that is tailored to the peculiar data access patterns of
backup applications and that strives for data durability.

For the sake of clarity, we now explain the operation of a baseline
P2P backup application. We gloss over the details of how data
redundancy is achieved and discuss the salient phases of the life-time
of backup data.

Using erasure coding, a backup object of size $o$ is encoded in $n$
\emph{fragments} of a fixed size $f$ which are ready to
be placed on remote peers. Any $k$ out of $n$ fragments are
sufficient to recover the original data;\footnote{For non-optimal
erasure-coding techniques such as Fountain Coding, as described in
Section~\ref{sec:coding}, this guarantee is given probabilistically.}
when using optimal erasure coding techniques, $k=\lceil o / f \rceil$.
The redundancy management mechanism determines the redundancy level $r
= nf / o$.

During the \textbf{backup phase}, data owners upload fragments to some
selected remote peers. We assume that any peer can collect a list of
remote peers with available storage space: this can be achieved with known techniques, \textit{e.g.} a central coordinator or a decentralized
data structure such as a distributed hash table. The backup phase
completes when \textit{all} $n$ fragments are placed on remote peers.

Once the backup phase is completed, the \textbf{maintenance phase}
begins. The purpose of this phase is to reestablish the desired
redundancy level in the system, that may decrease due to peer deaths:
new fragments must be re-injected in the system. The crux of
data maintenance is to determine when the redundancy of the backup
object is too low to allow data recovery and to generate other
fragments to rebalance it. In the event of a peer death, the
system may trigger the maintenance phase immediately (eager repairs)
or may wait for a number of fragments to be tagged as lost
before proceeding with the repairs (lazy repairs)
\cite{Kiran04totalrecall, dimakis07, duminuco-biersack-08}. As such,
it is important to discern \textit{unambiguously} permanent deaths
from the normal online behavior of peers: this is generally achieved
by setting a time-out value, $\Theta$, for long-term peer
unavailability.

Note that, as peers hold a local copy of their data, maintenance can
be executed solely by the data owner, or (as often done in storage
systems) it can be delegated. In both cases, it is important to
consider the timeframe in which data cannot be maintained. First,
fragments may be lost before a host failure is detected
using the time-out mechanism outlined above. This problem is
exacerbated by the availability pattern of the entity (data owner or
other peers) in charge of the maintenance operation: indeed, host
failures cannot be detected during the offline periods. Second, data
loss can occur during the restore process. For this reason, in
Section~\ref{sec:redundancy}, we consider a redundancy management policy
that ensures data is not lost in the time-window $w = \Theta +
a_{\mathrm{off}}$, where $a_{\mathrm{off}}$ is the (largest) transient off-line period
of the entity in charge of data maintenance. For example, if the data
owner executes data maintenance: first, it needs to be on-line to
generate new fragments and upload them, and second, the
timeout $\Theta$ has to be expired. Additionally, our mechanism
selects a redundancy level such that data loss does not occur before
the restore process is completed.

In the unfortunate case of a disk or host crash, the \textbf{restore
  phase} takes place. Data owners contact the remote machines holding
their fragments, download at least $k$ of them, and
reconstruct the original backup data.

Before proceeding, we now define the performance metrics we are
interested in for this work. Overall, we compute the performance of a
P2P backup application in terms of the amount of time required to
complete the backup and the restore phases, labelled \emph{time to
  backup} (TTB) and \emph{time to restore} (TTR). Moreover, in the
following sections, we use baseline values for backup and restore
operations which bound both TTB and TTR. We compute such bounds as
follows: let us assume an \textit{ideal} storage system with unlimited
capacity and uninterrupted online time that backs up user data. In
this case, TTB and TTR only depend on the size of a backup object and
on uplink bandwidth and availability of the data owner. We label these
ideal values \emph{minTTB} and \emph{minTTR}. Formally, we have that
a peer $i$ with upload and download bandwidth $u_i$ and $d_i$, starting
the backup of an object of size $o$ at time $t$, completes its backup
at time $t'$, after having spent $\frac o {u_i}$ time online.
Analogously, $i$ restores a backup object with the same size at $t''$
after having spent $\frac o {d_i}$ time online. Hence, we have that
$$minTTB(i, t)=t' - t$$ and $$minTTR(i, t)=t'' - t.$$ We use these
reference values throughout the paper to compare the relative
performance of our P2P application versus that of such an ideal
system.

\section{Redundancy Management}
\label{sec:redundancy}
We now discuss the key idea of our work: a redundancy management
mechanism to achieve data durability. In practice, data can be
considered as durable if the probability to lose it, due to the
permanent failure of hosts in the system, is negligible. The problem of
designing a system that guarantees data durability can be approached
under different angles.

As noted in previous works \cite{duminuco2007proactive, 1774229}, data
availability implies data durability: a system that injects sufficient
redundancy for data to be available at any time, coupled with
maintenance mechanisms, automatically achieves data durability. These
solutions are, however, too expensive in our scenario: the amount of
redundancy needed to guarantee availability is much higher than what
needed to obtain durability.

Instead of using high redundancy, data durability can also be achieved
with efficient maintenance techniques. For example, in a datacenter,
each host is continuously monitored: based on statistics such as the
mean time to failure of machines and their components, it is possible
to store data with very little redundancy and rely on system
monitoring to detect and react immediately to host failures. Failed
machines are replaced and data is rapidly repaired due to the
dedicated and over-dimensioned nature of datacenter networks.
Unfortunately, this approach is not feasible in a P2P setting. First,
the interplay of transient and permanent failures makes failure
detection a difficult task. Since it is difficult to discern deaths
from the ordinary online behavior of peers, the detection of permanent
failures requires a delay during which data may be lost. Furthermore,
data maintenance is not immediate: in a P2P application deployed on
the Internet, bandwidth scarceness and peer churn make the repair
operation slow.

In summary: on the one hand durability could be achieved with high
data redundancy, but the cost in terms of resources required by peers
would be overwhelming. On the other hand, with little redundancy,
durability could be achieved with timely detection of host failures
and fast repairs, which are not realistic in a P2P setting. 

The endeavor of this work is to design a redundancy management
mechanism that achieves data durability without requiring high
redundancy levels nor fast failure detection and repair mechanisms.
Our solution to the problem stems from the particular data access
workload of backup applications: data is written once, during backup,
and read (hopefully) rarely, during restores. Hence, we design a
mechanism that injects only the data redundancy level required to
compensate failure detection and data repair delays. That is, we
define data durability as follows.

\begin{definition}
 \textit{Data durability} $d$ is the probability to be able to access data
 after a \textit{time window} $t$, during which no maintenance
 operations can be executed.
\end{definition}

\begin{definition}
  The time window $t$ is defined as $t=w+TTR$, where $w$ accounts for
  failure detection delays and $TTR$ is the time required to download
  a number of fragments sufficient to recover the original data.
\end{definition}

As discussed in Section~\ref{sec:motivations}, $w$ depends on whether the
maintenance is executed by the data owner or is delegated, and can be
thought of a parameter of our scheme. 

The goal of our redundancy management mechanism is to determine the
data redundancy that achieves a target data durability: we proceed as
follows. A peer with $n$ fragments placed on remote peers
could lose its data if more than $n - k$ of them would get lost as
well within the time window $t$. The data redundancy required to avoid
this event is $r=n/k$.  

Peer deaths can be determined by disks and host crashes, or by human
events such as users uninstalling the application and leaving the
network. Disk drives in practical settings have a lifetime of several
years on average~\cite{fast07}; in the evaluation section, we will
evaluate our strategies in challenging sitiations where peer lifetime
ranges between a few months and a few years. Let us assume peer deaths
to be memoryless events, with constant probability for any peer and at
any time. Peer lifetimes are exponentially distributed stochastic
variables with a parametric average $\tau$. Hence, the probability for
a peer to be alive after a time $t$ is $ e^{-t/\tau}$. Assuming death
events are independent, data durability writes as:

\begin{equation}
d = \sum_{i=k}^{n}{n \choose
  i} \left( e^{-t/\tau}\right)^{i}\left(1 - e^{-t/\tau}\right)^{n-i}.
\label{eq:durability}
\end{equation}

Equation~\ref{eq:durability} depends on $t$ which, in turn, is a function
of TTR. However, peers cannot readily compute their TTR, as this
quantity depends on the characteristics of remote peers hosting their
fragments. We thus propose to use the following heuristic as a method
to \textit{estimate} the TTR. Suppose peer $p_0$ is computing an
estimate of its TTR. In the event of a crash, we assume $p_0$ to
remain online during the whole restore process. In such a case,
assuming no network bottlenecks, its TTR can be bounded for two
reasons: 
\begin{enumerate}
\item the download bandwidth $D_0$ of peer $p_0$ is the
bottleneck;
\item the upload rate of remote peers holding $p_0$'s
data is the bottleneck.
\end{enumerate}
Let us focus on the second case: we define the
\emph{expected upload rate} $\mu_i$ of a generic remote peer $p_i$
holding a backup fragment of $p_0$ as the product of the availability
of peer $p_i$ and its upload bandwidth, that is $\mu_i=u_i a_i$.

Peer $p_0$ needs to download at least $k$ fragments to fully recover a
backup object. Let us assume these $k$ fragments are served by the $k$
remote peers with the highest expected upload rate $\mu_i$. In this
case, the ``bottleneck'' is the $k$-th peer with the lowest expected
upload rate $\mu_k$. Then, an estimation of TTR, that we label \emph{eTTR},
can be obtained as follows:

\begin{equation}
eTTR = 
\max \left(\frac o {D_0}, \frac o {k\mu_k} \right).
\label{eq:ettr}
\end{equation}

We now set off to describe how our redundancy management scheme works
in practice: the redundancy level applied to backup data is computed
by the combination of Equation~\ref{eq:durability} and Equation~\ref{eq:ettr}.
Let us assume, for the sake of simplicity, the presence of a central
coordinator that performs membership management of the P2P network:
the coordinator keeps track of users subscribed to the application,
along with short-term measurements of their availability, their
(application-level) uplink capacity and the average death rate $\tau$
in the system. While a decentralized approach to membership management
and system monitoring is an appealing research subject, it is common
practice (\textit{e.g.}, Wuala\footnote{\url{http://www.wuala.com}})
to rely on a centralized infrastructure and a simple heartbeat
mechanism.

During a backup operation, peers query the coordinator to obtain
remote hosts that can be used to store fragments, along with their
availability. A peer constructs a backup object, and subsequently
uploads $k$ fragments to distinct, randomly selected available remote
hosts. Then the peer continues to inject redundancy in the system, by
sending additional fragments to randomly selected available peers,
until a stop condition is met. Every time one (or more) new fragment
is uploaded, the peer computes $d$ and eTTR: the stop condition is
met if $d \geq \sigma_1$ and $eTTR \leq
\sigma_2$. While selecting an appropriate $\sigma_1$ is trivial, in
the following we define $\sigma_2$ as $\sigma_2 = \alpha \cdot
minTTR$, where $\alpha$ is a parameter that specifies the degradation
of TTR with respect to an ideal system, tolerated by users.

We now discuss in details the influence of the two stop conditions on
the behavior of our mechanism. Given Equation~\ref{eq:durability}, we study
the impact of the ratio $\frac{w+eTTR}{\tau}$:
\begin{itemize}
\item $\tau \gg w+eTTR$: this case is representative of a ``mature''
  P2P application in which the dominant factor that characterizes peer
  deaths are permanent host failures, rather than users abandoning
  the system. Hence, the exponential in Equation~\ref{eq:durability}
  is close to 1, which implies that the target durability $\sigma_1$ can be
  achieved with a small $n$.

  As such, the condition on $eTTR \leq \sigma_2$ prevails on $d \geq
  \sigma_1$ in determining the redundancy level to apply to backup
  data. This means that the accuracy of the estimate eTTR plays an
  important role in guaranteeing acceptable restore times; instead,
  errors on eTTR have no impact on data durability.

\item $\tau \sim w+eTTR$: this case is representative of a P2P
  application in the early stages of its deployment, where the abandon
  rate of users is crucial in determining the death rate. In this
  case, the exponential in Equation~\ref{eq:durability} can be arbitrarily
  small, which implies that $n \gg k$, \textit{i.e.,} the target
  durability $d$ requires higher data redundancy.

  In this case, the condition $d \geq \sigma_1$ prevails on $eTTR \leq
  \sigma_2$. Hence, estimation errors on the restore times may have an
  impact on data durability: \textit{e.g.}, underestimating the TTR
  may cause $n$ to be too small to guarantee the target $\sigma_1$. In
  Section~\ref{sec:evaluation}, we study this scenario.
\end{itemize}

In summary, the key idea of our redundancy management mechanism is
that the redundancy level applied to backup data is computed in an
\textit{on-line} manner, during the backup phase. This comes in sharp
contrast to computing the redundancy level in an \textit{off-line}
manner, solely based on system-wide statistics, that characterize
previous approaches to redundancy management. 

A by-product of our approach is that our mechanism can \textit{adapt}
the redundancy rate $r$ each peer applies to its data based on system
dynamics. Now, we must prove that the system reaches a \textit{stable
  state}: system dynamics must not bring the redundancy mechanism to
oscillate around $r$. Based on Equation~\ref{eq:durability} and
Equation~\ref{eq:ettr}, we face a retroactive system in which a feedback
loop exists on the durability $d$. Given a target durability $d$, a
system-wide average death rate $\tau$ and a time window $t = w +
eTTR$, we can derive $r$.  The problem is that eTTR depends on the
short-term behavior of peers as well as the redundancy rate $r$.

First, we study how eTTR and $d$ vary as a function of the redundancy
rate $r$.

\begin{proposition}
\label{proposition:ettr}
  $eTTR$ is a non-increasing function in $r$.
\end{proposition}
\noindent \textit{Sketch of the proof:} Recall that $r=\frac{n
  f}{o}$. Let us assume a peer $p_0$ has the following ranked list of
remote peers: $$\{ \mu_1, \mu_2, \mu_3, ..., \mu_k \},$$ where, without
loss of generality, $\mu_i < \mu_j\, \forall i<j$. If $r$ increases,
then $n$ increases: new fragments must be stored on new remote
peers. For simplicity, assume a single fragment is to be placed on
peer $p_q$.

Two cases can happen:
\begin{enumerate}
  \item $\mu_q > \mu_k$; in this case, eTTR remains unvaried, since
$p_q$ is ``slower'' than the $k$-th peer used to compute eTTR; \item
$\mu_q < \mu_k$; in this case, $p_q$ ``ejects'' the current $k$-th
peer from the ranked list defined above.
\end{enumerate}

As such, eTTR can only decrease. Note that eTTR may not reach
the stop condition $\sigma_2$ if the parameter $\alpha$ is not
appropriately chosen: simply stated, a \textit{plateau} value of eTTR
exists when placing fragments on all peers in the network.

\begin{proposition}
\label{proposition:durability}
  $d$ is an increasing function in $r$.
\end{proposition}
\noindent \textit{Sketch of the proof:} Equation~\ref{eq:durability} is a
composite function of eTTR. Hence, by increasing $r$, new fragments
have to be placed on remote peers and it is not guaranteed, in
general, that this contributes to decrease $d$.  However, thanks to
Proposition~\ref{proposition:ettr}, eTTR is non-decreasing in $r$,
hence $t=w+eTTR$ is non decreasing in $r$. As a consequence, $d$ is an
increasing function in $r$.

We can now state the following Proposition:
\begin{proposition}
\label{proposition:stability}
  The redundancy management mechanism presented in this section is
  stable.
\end{proposition}
\noindent \textit{Sketch of the proof:} By design, our redundancy
mechanism shall only increase $r$. Now,
Proposition~\ref{proposition:ettr} states that increasing $r$ yields
lower values of eTTR, hence, eventually, the system either arrives
at the stop condition $eTTR \leq \sigma_2$, when $\alpha$ is chosen
appropriately, or it reaches the plateau defined above. Similarly, by
Proposition~\ref{proposition:durability}, increasing the redundancy in
the system implies that $d$ grows asymptotically to 1, hence the
system eventually reaches the stop condition $d \geq \sigma_1$.\\

It is natural to question why in
Proposition~\ref{proposition:stability} we omit the possibility of
removing fragments from remote peers if $r$ is too high. Let us
consider such an operation: one possibility would be to drop a remote
fragment at random. This operation would be unstable: indeed, for
example, deleting a fragment from the ``fastest'' peer in the ranked
list defined above would increase eTTR, decrease $d$, which as a
consequence might require to re-inject a fragment. Instead, we could
delete fragments starting from the ``slowest'' peer: in this case, the
drop operation would be stable, but the storage load in the system may
eventually become concentrated on fast peers only. Moreover, avoiding
deletions can spare maintenance operations in the future should one or
more of the remaining fragments on remote peers be lost. Due to these
reasons, in this work we do not allow fragments to be dropped.

\section{Coding and Data Management}
\label{sec:coding}
With the redundancy management mechanism described in
Section~\ref{sec:redundancy}, the redundancy level applied to backup data
is computed in an \textit{on-line} manner. Instead, the redundancy
rate used in most related work is usually computed \textit{off-line},
given sufficiently representative statistics on the system, including
transient and non-transient failures. These system-wide statistics are
used to compute a unique redundancy rate that every peer will
use. Instead, our approach requires each peer to compute
an \emph{individual} redundancy level: the time window $t$ is a
function of $eTTR$, which is different for every peer.

The endeavor of this Section is to study the practical implications
that stem from adopting two families of coding algorithms to implement
our redundancy management mechanism. 
Indeed, linear coding techniques assume data fragments to be grouped
in \emph{blocks} and each block to be processed separately to create
redundancy. The block size depends on the application requirements and
on the coding technique used. In the context of incremental backup,
the high number of initial fragments, the always increasing amount of
data and its generation rate make the block size play an important
role in the system design. Given the complexity of the topic and the
space constraints, in this paper we will consider the backup object to
be composed of different coding blocks of the same size.

We remark that we have implemented a library in C (with Python
wrappers) of the coding technique described in
Section~\ref{sec:fountain}. This library is a core component of a
prototype P2P backup application that we are currently testing in a
synthetic environment. 

\subsection{Optimal Erasure Coding}
Erasure coding introduces data redundancy by transforming an original
file composed of fragments into a longer file such that the original
file can be recovered from a subset of the encoded fragments. More
formally, assuming the backup object to be segmented in blocks of $k$
fragments each, each portion of the original data will be recovered if
a sufficient number of the $n$ encoded fragments will be successfully
received.  An erasure code is optimal if any $k$ out of the $n$
encoded fragments are sufficient to recover the original block. The
\textit{code rate}\footnote{In this work, code rate and redundancy
  rate are used as synonyms.} is defined as $r=n/k$ and represents the
number of ``redundant'' fragments per ``useful'' fragments generated
by the encoder. Note that optimal codes are often costly when $n$ is
large: practical solutions usually have \textit{quadratic} encoding
and decoding complexity.

Among the optimal techniques, Reed-Solomon (RS) codes are the most
widely used. A RS code is defined by the couple of parameters $k$ and
$n$, which are fixed \emph{a priori} and and are
systematic, \textit{i.e.}, the RS encoded data consists of $k$
original fragments and $n-k$ encoded fragments.

There are several implications due to the lack of flexibility of RS
codes. First, the coding rate must be computed off-line, which poses a
fundamental challenge to use them as a base for our redundancy
mechanism. Furthermore, any change in the redundancy level resulting
from system dynamics, entails a complete re-initialization of the
encoder: thus, coding/decoding operations are strictly dependent on
the specific configuration of the RS scheme. Additional complications
arise when considering repair operations. Since each RS encoded
fragment is unique, the entity in charge of data maintenance (the data
owner in our case) must maintain metadata information concerning which
remote peer store which encoded fragment.

In case a repair is needed,
the data owner must repair exactly the lost encoded fragment. In the
following, we will outline several techniques that can be used to
overcome the limitations of RS codes when applied to our setting.

The problem of achieving a redundancy level in the system that is
determined in an on-line manner can be addressed simply as follows.
The encoder can be parametrized by ``overshooting'' the coding rate:
as such a $(k,n)$ RS-code becomes a $(k,n')$ RS-code with $n' > n$.
Note that the RS coding matrix discussed above need not be generated
all at once: additional rows can be computed when required.  In
practice, with reference to the adaptive mechanism discussed in
Section~\ref{sec:redundancy}, we can choose a large $n'$, place $k$
encoded fragments on remote hosts, then incrementally inject
redundancy in the system until the stop condition is met.
Furthermore, as long as $n'$ is sufficiently high, our redundancy
management mechanism can inject more redundancy in the system, or drop
excess fragments, if required. Assuming that a value of $n'$ large
enough can be found, the price to pay is a substantial bookkeeping
effort to store metadata information about which encoded fragments are
available in the system.

The technique described above allows to overcome the limitations on
the number of encoded fragments to be generated, but does not solve
the limitation on the source size. In fact, the number of fragments to
backup can vary, which suggest to break the backup object into several
coding blocks. This allows to encode separately blocks of data and to
tune their redundancy independently.

This approach is convenient when a certain encoded portion of the
backup object has low redundancy. In this case, only the blocks
associated to that portion of data will be considered for the
generation of the new encoded fragments, resulting in an improvement
of coding speed (being the block size smaller than the one that would
contain the whole backup object). The drawback of this approach is
that enough fragments of every encoding block have to be received.  As
a consequence, this approach is prone to burst loss -- that is, the
loss of several fragments beloning to the same encoding block -- which
could make the recovery phase to fail.

The burst loss problem can be addressed using a well known technique in
digital communications: \textit{block interleaving}. 
In words, interleaving amounts to defining two levels of granularity to
partition backup data. The boundary of a \textit{backup block} is
defined such that for each block, only few fragments $k$ are required.
Moreover, with block interleaving, the parameters that govern the RS
coding scheme can be independent of the size of a backup
object.

Although interleaving techniques help mitigating the limitations of RS
codes, their efficient use presents non trivial challenges. In fact,
in order to minimize the padding, RS blocks of different sizes can be
used.  Unfortunately this means that the erasure protection capability
will be different for different segments of the backup object. Another
common problem of using interleaving techniques is the memory
consumption.
This means that the received fragments have to be processed all
together, thus increasing the processing time and the memory
requirements.

\subsection{Fountain Codes}
\label{sec:fountain}
Fountain Codes have been vastly studied in the literature, with
applications to digital communications \cite{byers2002digital},
content delivery \cite{byers2002informed}, storage
\cite{dimakis2006distributed} and P2P \cite{cataldi2009corp}.
This family of codes is particularly suitable to our goals because of
its unique characteristics. 
The generation of an encoded fragment is independent from the others
(\textit{on-the-fly} property) and the number of encoded fragments
that can be generated from the original data is potentially infinite
(\textit{rateless} property).

Fountain Codes are not optimal, in the sense that the number of
encoded fragments necessary to recover the original data is slightly
larger than the original number of fragments. This inefficiency
depends on the parameters of the coding technique and on the block
size\footnote{In the context of Fountain Codes, the encoding block is
defined exclusively by the number of fragments $k$, $n$ not being
defined \textit{a priori}.} and is negligible for large data
blocks. In practice, the loss of efficiency is acceptable, if one
considers the increased computational efficiency (even linear with the
block size) of this family of codes with respect to RS codes
(typically quadratic).

In the context of this work, Fountain Codes are very simple to
use in practice. Indeed, the information about the fragment generation
should be shared between the encoder and the decoder.\footnote{This
  information can be transmitted together with the encoded fragment,
  or the choice of the degree distribution and the random generator
  can be shared.} Instead, in our application, the encoder and the
decoder coexist in the same entity: the data owner. Hence, such
information needs not a complex infrastructure to be set up between
separate communicating parties, but can be simply treated as
``metadata'' information to be stored locally (and eventually backed
up).

Fountain Codes make the mechanism described in
Section~\ref{sec:redundancy} trivial to achieve: as long as the
conditions on the \textit{eTTR} and $d$ are not met, the encoder
continues to generate new unique encoded fragments on the fly. When
the stop condition is reached, the encoding process terminates. In
case system dynamics trigger the generation of new encoded fragments
(\textit{e.g.} because host availability decreases), these can be
simply generated as needed, with the same procedure described
above. If the redundancy level in the system exceeds what is required
(\textit{e.g.} because host availability increases), any encoded fragment can
be deleted.

  Fountain encoded fragments are statistically
``interchangeable'': any encoded fragment can be used to reconstruct
the original data and any encoded fragment can be replaced by any
newly generated encoded fragments. As a consequence, also maintenance
operations are simplified as peers need not track of the exact
encoded fragment to replace.

Another appealing characteristic of Fountain Codes is that, unlike RS
codes, the block size is not mathematically constrained. Nevertheless,
a solution based on Fountain Codes is not exempt from data management
problems. While these codes allow maximum flexibility, the
definition of the block size is a tradeoff between the coding
inefficiency (which suggests to use large blocks) and the number of
operations required for either encoding or decoding (even if
linear-time implementations exist, memory and delay considerations
suggest to use shorter blocks). This means that given a potentially
large backup object, what is the best strategy, encoding the whole
data thus minimizing the coding inefficiency, or segmenting into
smaller blocks decreasing the complexity?

We argue that the data object should be partitioned in several blocks
whose size should depend not only on coding complexity and
inefficiency, but also on the user data generation rate.  One of the
coding strategies that can increase the performance of the code whilst
maintaining shorter block size is the sliding-windowing
approach~\cite{bogino2007sliding, cataldi2010sliding}. This approach
virtually increases the encoding block by allowing the overlap of two
or more subsequent coding blocks (referred as ``windows''). The block
overlap is a design parameter that impacts the performance of the code
and its value can be decided \textit{a priori} or according to
customized coding strategies. The typical drawback of using the
sliding-windowing approach is an increase in the decoding delay and
memory consumption. However, for the considered application, if the
block size is moderate, their impacts are acceptable.

\section{Performance Evaluation}
\label{sec:evaluation}
In the following, we proceed with a trace-driven system simulation,
and focus on the performance metrics outlined in
Section~\ref{sec:motivations}. That is, we are interested in studying the
time required to backup and restore user data: we perform a
comparative study of the results achieved by a system using our
redundancy management scheme and the traditional approach used for
storage applications. For the latter case, we implement a technique in
which the coding rate is set once and for all based on a system-wide
average of host availability.

Note that, for the purpose of our study, it is not necessary to
implement in detail the coding mechanisms described in
Section~\ref{sec:coding}. All we need to know for the evaluation of
transfer times is the number of fragments each peer has to upload
during the backup operation.

We use traces as input to our simulator that cover both the online
behavior of peers and their uplink and downlink capacities. Instead,
long-term failures and the events of peers abandoning the
applications, which constitute the peer deaths, follow a simple model
driven by the parameter $\tau$, as explained in
Section~\ref{sec:redundancy}. Due to the lack of traces that represent
the realistic ``data production rate'' of Internet users, in this
simulation study we confine our attention to a homogeneous setting:
each user has an individual backup object of the same size.

\subsection{Datasets}\label{sec:datasets}

\begin{figure}
\centering
\includegraphics[width=\columnwidth]{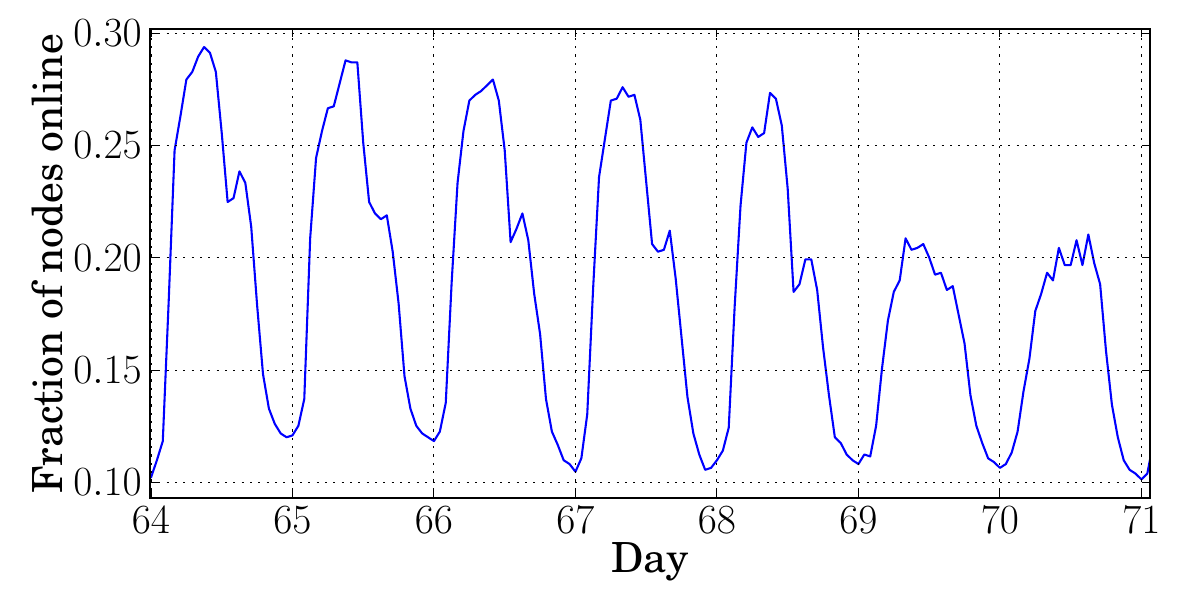}
\caption{Ratio of connected nodes in a representative week of the availability
trace. More users are connected during days than during night; in weekends, the total number of connected users drops.}
\label{fig:im-week}
\end{figure}

\paragraph*{\textbf{Availability trace}} The \emph{online behavior} of users,
i.e., their patterns of connection and disconnection over time, is
difficult to capture analytically. In this work we simulate a backup
application using a real application trace that exhibits both
heterogeneity and correlated user behavior. Our traces capture user
availability, in terms of login/logoff events, from an instant
messaging (IM) server for a duration of roughly 3 months. We argue
that the behavior of regular IM users constitutes a representative
case study. Indeed, for both an IM and an online backup application,
users are generally signed in for as long as their machine is
connected to the Internet; as it can be gleaned from
Figure~\vref{fig:im-week}, in this dataset it is possible to observe
strong diurnal and weekly patterns. Moreover, users have heterogeneous
behavior -- for example, some users often stay connected during
workdays while others have a less predictable uptime~\cite{bttf}.

\begin{figure}
  \centering
  \includegraphics[width=.4\textwidth]{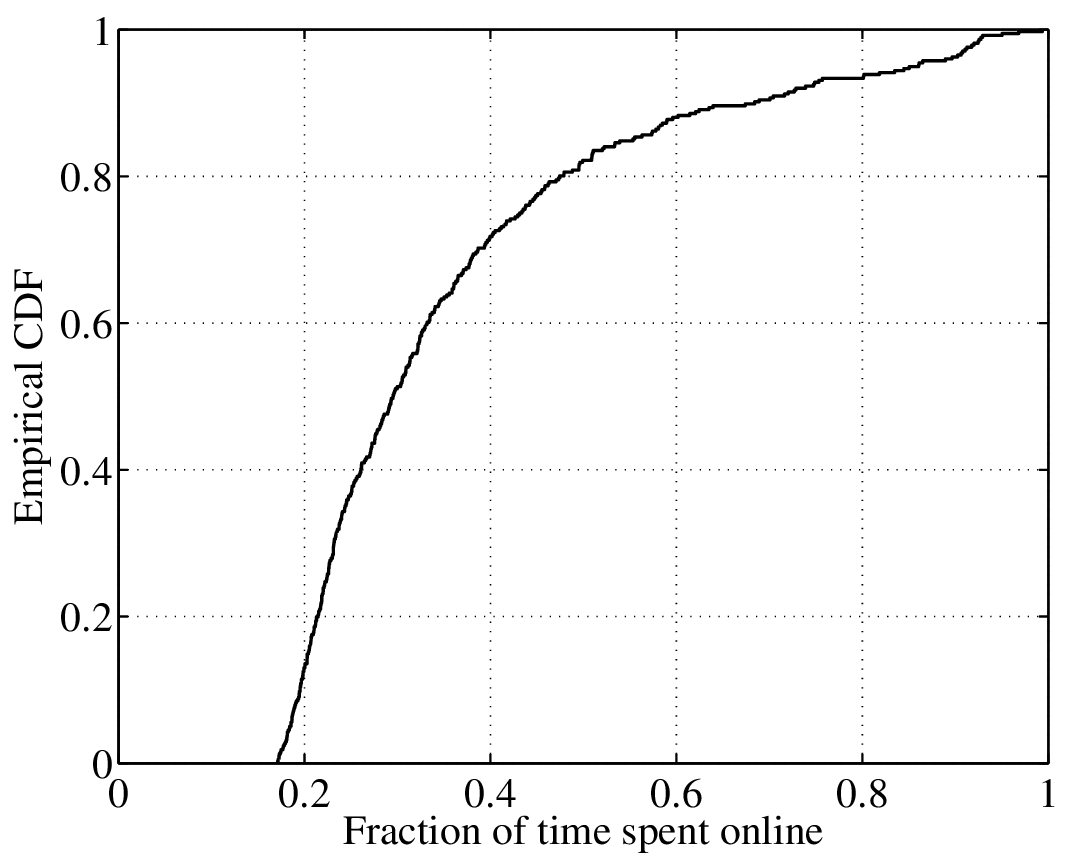}
\caption{CDF of the host availability from our traces. Note that users
spending less than 4 hours per day online are filtered from our data.}
\label{fig:traces}
\end{figure}

In this work we only consider users that are online for an average of
at least four hours per day, as done in Wuala~\cite{mager09}. Once
this filter is applied, we obtain the trace of 376 users. Since in P2P
storage systems the number of neighbors each node interacts with is
very often limited by design and scalability issues~\cite{p2p11}, we
believe this trace size is acceptable.  As shown in
Figure~\ref{fig:traces}, most users are online for less than 40\% of the
trace length, while some of them are almost always connected.

\paragraph*{\textbf{Bandwidth distribution}} Uplink capacities of peers are
obtained by sampling a real bandwidth distribution measured at more
than 300,000 unique Internet hosts for a 48 hour period from roughly
3,500 distinct ASes across 160 countries
\cite{Piatek07doincentives}. These values have a highly skewed
distribution, with a median of 77 KBps and a mean of 428 KBps. To
represent typical asymmetric residential Internet lines, we assign to
each peer a downlink speed equal to four times its uplink.

\subsection{Simulation Settings}
The trace-driven online behavior of a peer is overridden only during
the restore phase: in this work we make the assumption that in such
case, a peer remains online for the whole duration of the restore
process.

In our study, each peer has $o=10$ GB of data to backup (as soon as
the simulation begins), and dedicates 50 GB of storage space to the
application. The high ratio between these two values lets us disregard
issues due to insufficient storage capacity and focus on the subjects
of our investigation. The fragment size is set to 160 MB, implying a
minimum of $k=64$ fragments needed for restores.

We define peers' lifetimes to be exponentially distributed random
variables with an expected value $\tau = \{90 \text{ days}, 1
\text{ year}, 4 \text{ years} \}$, conservative values that are
noticeably lower than the disk failure rates measured in real-world
scenarios~\cite{fast07} (see Section~\ref{sec:redundancy}).  Besides peer
deaths, we study the impact of the $w$ parameter, which contributes to
the duration of the time-window for which our redundancy management
policy guarantees data durability, without maintenance (see
Section~\ref{sec:redundancy}).  As a reminder (see
Section~\ref{sec:motivations}), $w$ accounts for failure detection
delays. In our experiments $w$ takes values from 0 to 4 weeks.

Our adaptive redundancy policy uses the following parameters: we set
the thresholds $\sigma_1 = 0.9999$, so that the durability $d \geq
\sigma_1$ and $\sigma_2 \leq \max\left(1~\mathrm{day}, 2 \cdot
  minTTR\right)$ so that the estimated TTR $eTTR \leq \sigma_2$. In
this work, we compare against a baseline redundancy policy that aims
to guarantee data availability~\cite{Kiran04totalrecall}, labeled here
as ``availability-based''. Here we set a target data availability of
$t=0.99$, and use the system-wide average availability $a=0.36$ as
computed from our availability traces. Hence, we obtain a value
$n=228$ and a redundancy rate $n / k = 3.56$.

For each set of parameters, the simulation results are obtained by
averaging ten simulation runs.

\subsection{Results}
\label{sec:results}

\begin{figure}
\centering
\includegraphics[width=.4\textwidth]{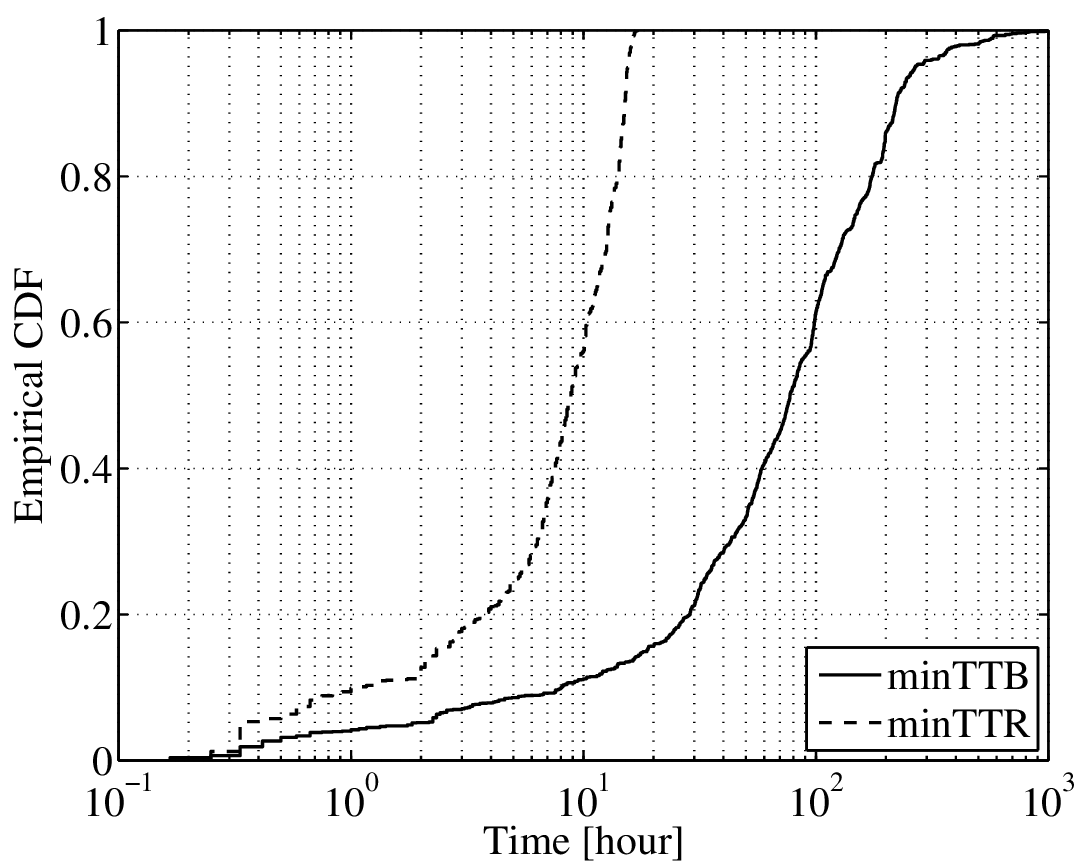}
\caption{CDF of the ideal times to transfer data: $minTTB$ and $minTTR$.}
\label{fig:minTTR}
\end{figure}

We begin our discussion by showing the bounds on TTB and TTR, as
defined in Section~\ref{sec:motivations}. Figure~\vref{fig:minTTR} shows
the cumulative distribution functions (CDF) of minTTB and minTTR
obtained using the input traces discussed above. Our working
assumption is that peers stay online during restore operations: as
such, only the (ordinary) backup phase suffers from peer
unavailability, and the distribution of minTTR depends only on the
bandwidth distribution, while minTTB also depends on the availability
traces.

While backup operations generally take days to complete, for a file
size of 10 GB, restore operations are several times faster. This can
be simply explained by the asymmetric bandwidth setup we use in our
simulations, and -- as discussed above -- by unavailability of peers
when data needs to be backed up. Since node bandwidth distribution is
skewed, a few nodes with very large bandwidth experience a much lower
value for both minTTR and minTTB; the tails with a very long minTTB
value are instead due to peers that remained disconnected for very
long time spans in our traces.

We now proceed to a detailed comparative study of our scheme to the traditional fixed-redundancy scheme. 
First, we focus on the data redundancy
level (that is, the code rate $r$) imposed by each approach.
\begin{figure}[bp]
  \centering
  \begin{tikzpicture}[font=\footnotesize,scale=0.9,tension=0.1]
    \pgfplotsset{every axis legend/.append style={
        cells={anchor=west}, at={(0.9,0.5)}, anchor=south,
        font=\scriptsize}}
    \pgfplotsset{every axis legend/.append style={font=\scriptsize}}
    \pgfplotsset{every axis plot/.append style={smooth}}
    \pgfplotsset{every axis/.append style={line width=0.5pt}}

    \begin{axis}[xlabel=$w$ (weeks),
      ylabel=Redundancy factor ($r$),
      grid=both]

     \addplot[color=black,mark=o,thick] file
     {./tikz/redundancy_0_0.txt};
     \addlegendentry{$eTTR$, $\tau=3$ months};
      \addplot[color=black,mark=x,thick] file
      {./tikz/redundancy_1_0.txt};
      \addlegendentry{$eTTR$, $\tau=1$ year};
      \addplot[color=black,mark=*,thick] file
      {./tikz/redundancy_2_0.txt};
      \addlegendentry{$eTTR$, $\tau=4$ years};

      \addplot[color=black,dashed] file
      {./tikz/redundancy_1_1.txt};
      \addlegendentry{availability-based};

   \end{axis}
 \end{tikzpicture}
 \caption{Redundancy rate as a function of $w$, for different values of $\tau$.}
 \label{fig:redundancy}
\end{figure}
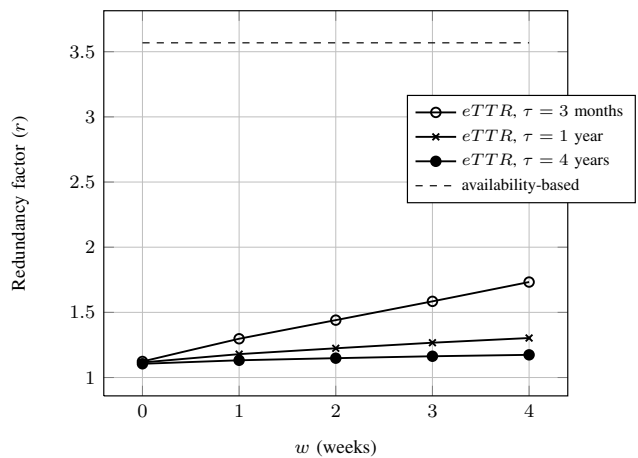

In Figure~\ref{fig:redundancy}, we show the average redundancy factor
for our mechanism and the one computed for the availability-based
scheme (which is fixed), as a function of the parameter $w$ and for
different values of $\tau$. We omit error bars from the plot as the
variance around the mean is negligible. Clearly, for increasing values
of $w$ the redundancy rate increases, as it is possible to evince from
Equation~\ref{eq:durability}. Note that our simulations account for a
realistic bandwidth distribution and for some real on-line user
behavior, which influence the eTTR
computation. Figure~\ref{fig:redundancy} also illustrates the impact of
$\tau$: when the dominant effect of non-transient failures is the
reliability of Internet hosts, that is $\tau$ is large, our mechanism
achieves data durability (and a controlled TTR) with a small
redundancy factor. Instead, when peer deaths are dominated by peers
abandoning the system, that is $\tau$ is small, our mechanism
compensates with a larger redundancy rate. In summary, our redundancy
management scheme obtains a redundancy factor ranging roughly between
\emph{half} and \emph{a third} of the availability-based scheme,
increasing the storage capacity of the system by a corresponding
factor between two and three. Since the amount of data to upload in
case of a disk crash is proportional to the redundancy level, the
impact of maintenance of system bandwidth decreases accordingly.

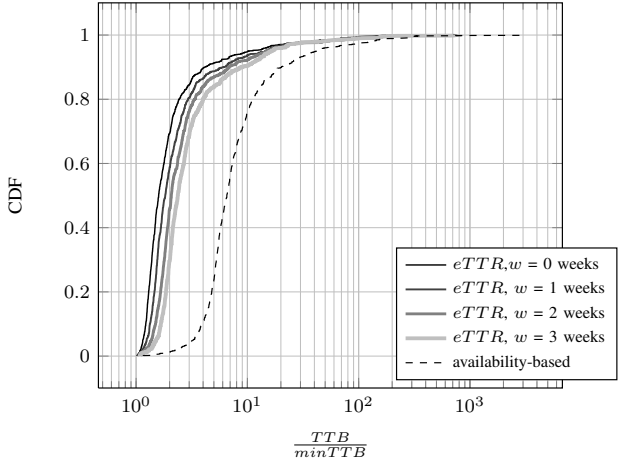
\begin{figure}[ht!]
  \centering
  \begin{tikzpicture}[font=\footnotesize,scale=0.9,tension=0.1]
    \pgfplotsset{every axis legend/.append style={
        cells={anchor=west}, at={(0.88,0.02)}, anchor=south,
        font=\scriptsize}}
    \pgfplotsset{every axis legend/.append style={font=\scriptsize}}
    \pgfplotsset{every axis plot/.append style={smooth}}
    \pgfplotsset{every axis/.append style={line width=0.5pt}}

    \begin{semilogxaxis}[xlabel=$\frac{TTB}{minTTB}$,
      ylabel=CDF,
      grid=both]

      \addplot[color=black, semithick] file
      {./tikz/cdf_TTB_0_0_0.txt};
      \addlegendentry{$eTTR$,$w$ = 0 weeks};

      \addplot[color=darkgray,thick] file
      {./tikz/cdf_TTB_0_1_0.txt};
      \addlegendentry{$eTTR$, $w$ = 1 weeks};

      \addplot[color=gray,very thick] file
      {./tikz/cdf_TTB_0_2_0.txt};
      \addlegendentry{$eTTR$, $w$ = 2 weeks};

      \addplot[color=lightgray,ultra thick] file
      {./tikz/cdf_TTB_0_3_0.txt};
      \addlegendentry{$eTTR$, $w$ = 3 weeks};

      \addplot[color=black, dashed] file
      {./tikz/cdf_TTB_0_0_1.txt};
      \addlegendentry{availability-based};

   \end{semilogxaxis}
 \end{tikzpicture}
 \caption{CDF of $TTB$ for different values of $w$, $\tau=3$ months.}
 \label{fig:ttb}
\end{figure}

\begin{figure}[ht!]
  \centering
  \begin{tikzpicture}[font=\footnotesize,scale=0.9,tension=0.1]
    \pgfplotsset{every axis legend/.append style={
        cells={anchor=west}, at={(0.88,0.02)}, anchor=south,
        font=\scriptsize}}
    \pgfplotsset{every axis legend/.append style={font=\scriptsize}}
    \pgfplotsset{every axis plot/.append style={smooth}}
    \pgfplotsset{every axis/.append style={line width=0.5pt}}

    \begin{semilogxaxis}[xlabel=$\frac{TTR}{minTTR}$,
      ylabel=CDF,
      grid=both]

      \addplot[color=black,semithick] file
      {./tikz/cdf_TTR_0_0_0.txt};
      \addlegendentry{$eTTR$, $w$ = 0 weeks};

     \addplot[color=darkgray,thick] file
      {./tikz/cdf_TTR_0_1_0.txt};
      \addlegendentry{$eTTR$, $w$ = 1 weeks};

      \addplot[color=gray,very thick] file
      {./tikz/cdf_TTR_0_2_0.txt};
      \addlegendentry{$eTTR$, $w$ = 2 weeks};

      \addplot[color=lightgray,ultra thick] file
      {./tikz/cdf_TTR_0_3_0.txt};
      \addlegendentry{$eTTR$, $w$ = 3 weeks};

      \addplot[color=black, dashed] file
      {./tikz/cdf_TTR_0_3_1.txt};
      \addlegendentry{availability-based};

   \end{semilogxaxis}
 \end{tikzpicture}
 \caption{Distribution of $TTR$ for different values of $w$, $\tau=3$ months.}
 \label{fig:ttr}
\end{figure}
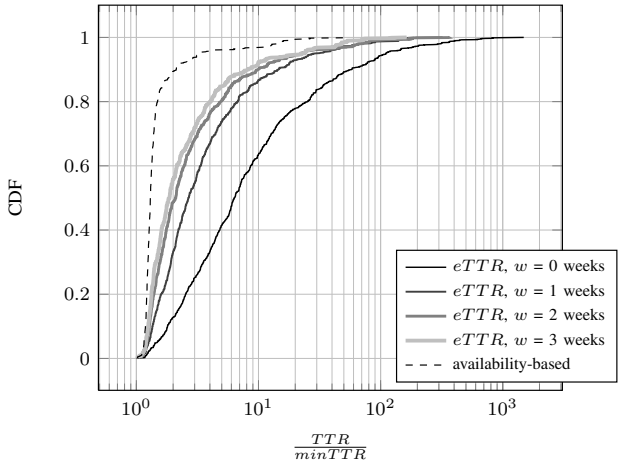

In addition to improving the aggregate storage capacity of the system,
our redundancy management scheme impacts both backup and restore
operations. Figure~\ref{fig:ttb} and \ref{fig:ttr} report the CDF of the
ratio of TTB and TTR over their respective ideal counterparts,
minTTB and minTTR. These plots are obtained with different values
of $w$, for a fixed $\tau=3$ months,\footnote{We present results for
$\tau=3$ months because the effects of $w$ are more marked. We obtain
similar qualitative results for larger values of $\tau$. Also, for
clarity of presentation, we omit the CDF for $w=4$ weeks.} and
illustrate the results of our mechanism and that achieved by the
availability-based scheme. Figure~\ref{fig:ttb} indicates that, due to a
lower redundancy factor, the median of the distribution of TTB is
roughly reduced by a factor of four. Moreover, increasing values of
$w$ have essentially little impact on TTB. The price to pay for fast
backup operations is shown in Figure~\ref{fig:ttr}: restore operations
take more time to complete w.r.t. a traditional approach to redundancy
management. Here the parameter $w$ plays an important role: for small
$w$ values, little redundancy is applied to backup data. As such, the
opportunity to retrieve enough encoded fragments to restore data is
largely affected by peer availability. Instead, when $w$ is large,
restore operations are more efficient and less sensitive to peer
availability.

In summary, our results support the rationale underlying the design of
our redundancy management scheme: TTB is generally several times
larger than TTR, even in an ideal case (as shown in
Figure~\ref{fig:minTTR}). Because of this unbalance, we argue that it is
reasonable to use a redundancy management scheme that trades longer
TTR (which affects only users that suffer a crash) for shorter TTB
(which affects all users).

Now, we delve into the details of our scheme and study its sensitivity
to errors due to the heuristic we use to estimate TTR. The main
reason for errors on eTTR are due to the fact that the heuristic
defined in Equation~\ref{eq:ettr} assumes $k$ encoded fragments to be
downloaded from the $k$ fastest peers that hold backup data. In
practice, however, the $k$ encoded fragments are downloaded from the
peers that are available when a restore operation is
executed. Depending on the bandwidth distribution of the peers in the
system, such difference can cause the estimated TTR value to be
different from what achieved in practice.

\begin{figure}[ht]
  \centering
  \begin{tikzpicture}[font=\footnotesize,scale=0.9,tension=0.1]
    \pgfplotsset{every axis legend/.append style={
        cells={anchor=west}, at={(0.25,0.9)}, anchor=north,
        font=\scriptsize}}
    \pgfplotsset{every axis legend/.append style={font=\scriptsize}}
    \pgfplotsset{every axis plot/.append style={smooth}}
    \pgfplotsset{every axis/.append style={line width=0.5pt}}

    \begin{axis}[xlabel=Estimation error: $\frac{|TTR - eTTR|}{TTR}$,
      ylabel=Durability error: $\frac{\sigma_1-d}{\sigma_1}$,
      grid=both]

      \addplot[color=black, mark=o,thick] file
      {./tikz/error_TTR_0.txt};
      \addlegendentry{$\tau = 3$ weeks};
      \addplot[color=black,mark=x, thick] file
      {./tikz/error_TTR_1.txt};
      \addlegendentry{$\tau = 1$ year};
      \addplot[color=black, mark=*, thick] file
      {./tikz/error_TTR_2.txt};
      \addlegendentry{$\tau = 4$ years};

   \end{axis}
 \end{tikzpicture}
 \caption{Correlation between estimation errors and data durability.}
 \label{fig:durability}
\end{figure}
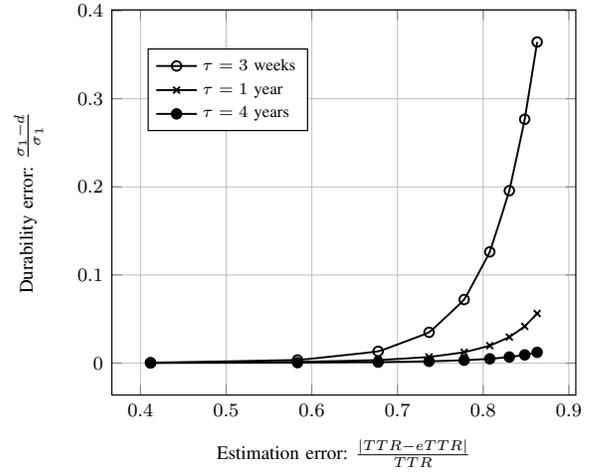

Now, if eTTR is larger than TTR, more redundant data is injected
in the system, which has no negative impact on data durability. What
is the impact on durability if peers underestimate the TTR?  Using
Equation~\ref{eq:durability}, we compute the redundancy factor $r$, as a
function of eTTR, that meets the traget durability $\sigma_1$. Then,
using TTR and $r$, we compute the data durability
$d$. Figure~\ref{fig:durability} shows the impact of the relative
estimation error -- from a precise evaluation of TTR up to an error
of almost twice the TTR -- on the relative durability error using
the procedure described above, for different values of $\tau$ and for
$w=2$ weeks.

When $\tau$ is large, we have that $w+eTTR \ll \tau$: as such, even
large estimation errors have little impact on the durability
$d$. Instead, when $\tau$ is small, we have that $w+eTTR \simeq \tau$: in
this case, data durability is more sensitive to estimation errors. As
a consequence, data redundancy may not be sufficient and data loss
events may occur.

In Table~\ref{tab:dataloss}, we illustrate the effects discussed by
quantifying data loss events for $w=2$ weeks. Here we count the
percentage of peers that have not been able to restore their data
after a local disk crash, averaged over 10 simulation runs. We break
down the data loss cases between \emph{incomplete backup}
and \emph{failed restore}: the latter case encompasses all cases where
peers lose data after completing their backup. Furthermore, we also
specify the percentage of \emph{unavoidable} cases in which peers fail
before $minTTB$: in this case, not even an ideal system could have
guaranteed a safe backup.

\begin{table}
\centering

\caption{Categorization of data loss events}
\label{tab:dataloss}
\begin{tabular}{|l||r||r|r|r|}
\hline
Avg. lifetime & Total & \multicolumn{2}{|c|}{Incomplete backup} &Failed \\
\cline{3-4}
($\tau$)         & events       & Total  & Unavoidable&restore \\
\hline
\hline

3 months & 13\% & 10.4\% & 8.4\%  & \textbf{2.6\%} \\

\hline

1 year & 2.6\%  & 2.6\% & 2.3\% & \textbf{None} \\

\hline

4 years & 0.5\%  & 0.5\% & 0.25\% & \textbf{None} \\

\hline

\end{tabular}
\end{table}

A lesson we can draw from Table~\ref{tab:dataloss} is that most data
loss episodes are simply due to node failure \emph{before the backup
is completed}; this result confirms that it is sensible to optimize
time to backup by reducing redundancy and hence also network load. In
addition, it can be noted that a large majority of data loss episodes
are \emph{unavoidable} with any online storage solution: nodes with
low bandwidth risk crashing before completing uploads even if saving
data to a reliable server with 100\% uptime and unlimited bandwidth.
``Failed restore'' events -- present only in unstable systems with low
$\tau$ -- are imputable to the impact of estimation error on
durability, at discussed above. However, we remark that the impact of
this effect even in such a situation is outnumbered by the unavoidable
data loss episodes; this leads us to conclude that nodes with very low
lifetime are intrinsically unsuited to any kind of online storage
solution, and not only to P2P backup.

\section{Related Work}
\label{sec:related}

Redundancy rates and data repair techniques in P2P backup systems have
been investigated from various angles. Earlier
works~\cite{batten2001pstore, cox02pastiche, 1021936} adopt simple
replication strategies, resulting in higher storage and bandwidth
costs for backing up and maintaining data in the system. In other
proposals, erasure coding is used in order to obtain high durability
while minimizing storage costs on nodes, but redundancy values are
fixed parameters that are chosen by the designer independently of the
system
characteristics~\cite{Sameh03acooperative,Haeberlen05glacier}. The
Wuala online storage service encodes data on peers with a fixed
redundancy level and avoids the need for maintenance by storing a full
replica of the data in central servers~\cite{mager09}.  More elaborate
policies belong to two different categories: in some
cases~\cite{Chun06efficientreplica, 1774229}, redundancy is determined
as a function of node failure rate in order to guarantee data
durability at the expense of data availability. Many other approaches
(\textit{e.g.}, \cite{Kubiatowicz00oceanstore, Kiran04totalrecall}) guarantee
low latency through prompt data availability, but require high
redundancy rates in typical settings. In contrast with these
approaches, our proposal strives to provide \emph{both} durability and
performance at a low redundancy cost, relaxing prompt data
availability by requiring that data becomes recoverable within a given
time window. Finally, Pamies-Juarez \emph{et
al.}~\cite{pamies2010availability} investigate the relationship
between redundancy and data retrieval times, but they center their
investigations on cases where the online session length duration is
orders of magnitude shorter than the length of a data transfer
process; this scenario is clearly not applicable to our case of long
restore processes.

A complete system design requires considering several problems that
were not addressed in this paper; fortunately, many of them have been
tackled in the literature.

When a full system needs to be backed up, \emph{convergent encryption}
\cite{cox02pastiche, 1021936} can be used to avoid backing up duplicate
copies of the same file owned by different users.

Data maintenance is cheap in our scenario, where it is performed by a
data owner with a local copy. When maintenance is delegated to nodes
that do not have a local copy of the backup objects, various coding
schemes can be used \cite{dimakis07,duminuco-biersack-08} to
limit the amount of required data transit. For these settings,
cryptographic protocols \cite{oualha2008security,
ateniese2008scalable} have been designed to verify the authenticity of
stored data.

A recurrent problem for P2P applications is guaranteeing load
balancing and creating incentives to encourage nodes in contributing
more resources. This can be done via reputation
systems \cite{kamvar2003eigentrust} or virtual
currency \cite{vishnumurthy2003karma}. Specifically for storage
systems, an easy and efficient solution is segregating nodes in
sub-networks with roughly homogeneous characteristics such as uptime
and storage space \cite{Pamies-juarez_rewardingstability,
EURECOM+2738}.

Backup objects, whose confidentiality can be ensured by standard
encryption techniques, should encode incremental differences between
archive versions. Various techniques have been proposed to optimize
computational time and size of these
differences \cite{Tangwongsan_efficientsimilarity}.

It may happen that resources offered by peers are just not sufficient
to satisfy all user needs. In this case, a hybrid peer-assisted system
can be developed where data is stored on a centralized data center and
on peers. This can result in scenarios having a performance comparable
to centralized systems, at a fraction of the costs
\cite{EURECOM+3140}.

\section{Conclusion}
\label{sec:conclusion}
In this work we focused on P2P backup systems, and designed a
redundancy management mechanism tailored to the specific data access
patterns that characterize data backup. The goal of our mechanism was
to achieve data durability without requiring large redundancy factors
(typical of storage applications) nor fast failure detection
mechanisms.

Our experiments showed that, in a realistic setting, a redundancy that
caters to data durability can be less than half of what is needed to
guarantee availability. This results in a system with a storage
capacity that is more than doubled, and backup operations that are
much faster (up to a factor of 4) than on a backup system based on
traditional redundancy management. This latter property is
particularly desirable since, in most of the cases, peers suffering
data loss were those that could not complete the backup before
crashing.

We also showed that the price to pay for efficient backup operations
was a decreased (but controlled) performance of restore operations. We
argued that this was a reasonable penalty, considering that all peers
in the system would benefit from backup efficiency, while only those
peers suffering from a failure would have to bear longer
restore times.

Finally, we studied data loss events: our results indicated that such
events are practically negligible for a mature P2P application in
which permanent host failures dominate peer deaths. We also
showed the limitations of our technique for a system characterized by
a high application-level churn, which is typical of new P2P applications that must
conquer user trust.

\bibliographystyle{plain}
\bibliography{references}

\begin{thebibliography}{10}
\providecommand{\url}[1]{#1}
\csname url@samestyle\endcsname
\providecommand{\newblock}{\relax}
\providecommand{\bibinfo}[2]{#2}
\providecommand{\BIBentrySTDinterwordspacing}{\spaceskip=0pt\relax}
\providecommand{\BIBentryALTinterwordstretchfactor}{4}
\providecommand{\BIBentryALTinterwordspacing}{\spaceskip=\fontdimen2\font plus
\BIBentryALTinterwordstretchfactor\fontdimen3\font minus
  \fontdimen4\font\relax}
\providecommand{\BIBforeignlanguage}[2]{{%
\expandafter\ifx\csname l@#1\endcsname\relax
\typeout{** WARNING: IEEEtran.bst: No hyphenation pattern has been}%
\typeout{** loaded for the language `#1'. Using the pattern for}%
\typeout{** the default language instead.}%
\else
\language=\csname l@#1\endcsname
\fi
#2}}
\providecommand{\BIBdecl}{\relax}
\BIBdecl

\bibitem{carbonite}
\BIBentryALTinterwordspacing
R.~Wauters. (2009) Online backup company carbonite loses customers' data,
  blames and sues suppliers. TechCrunch. [Online]. Available:
  \url{http://tcrn.ch/dABxRn}
\BIBentrySTDinterwordspacing

\bibitem{dropbox}
\BIBentryALTinterwordspacing
A.~Ferdowsi. (2011) Yesterday's authentication bug. Blog post. [Online].
  Available: \url{http://blog.dropbox.com/?p=821}
\BIBentrySTDinterwordspacing

\bibitem{dropio}
\BIBentryALTinterwordspacing
drop.io. (2010) An important update on the future of drop.io. Blog post.
  [Online]. Available:
  \url{http://http://blog.drop.io/2010/10/29/an-important-update-on-the-future-of-drop-io/}
\BIBentrySTDinterwordspacing

\bibitem{blake-rodrigues03}
C.~Blake and R.~Rodrigues, ``High availability, scalable storage, dynamic peer
  networks: pick two,'' in \emph{USENIX HotOS}, 2003.

\bibitem{Kiran04totalrecall}
R.~Bhagwan, K.~Tati, Y.~chung Cheng, S.~Savage, and G.~M. Voelker, ``Total
  recall: System support for automated availability management,'' in
  \emph{USENIX NSDI}, 2004.

\bibitem{farsite}
A.~Adya, W.~Bolosky, M.~Castro, G.~Cermak, R.~Chaiken, J.~Douceur, J.~Howell,
  J.~Lorch, M.~Theimer, and R.~Wattenhofer, ``{FARSITE: Federated, available,
  and reliable storage for an incompletely trusted environment},'' \emph{ACM
  SIGOPS Operating Systems Review}, vol.~36, 2002.

\bibitem{dimakis07}
A.~G. Dimakis, P.~B. Godfrey, M.~J. Wainwright, and K.~Ramchandran, ``Network
  coding for distributed storage systems,'' in \emph{IEEE INFOCOM}, 2007.

\bibitem{duminuco-biersack-08}
A.~Duminuco and E.~Biersack, ``Hierarchical codes: How to make erasure codes
  attractive for peer-to-peer storage systems,'' in \emph{IEEE P2P}, 2008.

\bibitem{duminuco2007proactive}
A.~Duminuco, E.~Biersack, and T.~En-Najjary, ``{Proactive replication in
  distributed storage systems using machine availability estimation},'' in
  \emph{ACM CoNEXT}, 2007.

\bibitem{1774229}
L.~Pamies-Juarez and P.~Garcia-Lopez, ``Maintaining data reliability without
  availability in p2p storage systems,'' in \emph{ACM SAC}, 2010.

\bibitem{fast07}
B.~Schroeder and G.~A. Gibson, ``Disk failures in the real world: What does an
  mttf of 1,000,000 hours mean to you?'' in \emph{USENIX FAST}, 2007.

\bibitem{byers2002digital}
J.~Byers, M.~Luby, and M.~Mitzenmacher, ``{A digital fountain approach to
  asynchronous reliable multicast},'' \emph{Selected Areas in Communications,
  IEEE Journal on}, vol.~20, no.~8, pp. 1528--1540, 2002.

\bibitem{byers2002informed}
J.~Byers, J.~Considine, M.~Mitzenmacher, and S.~Rost, ``{Informed content
  delivery across adaptive overlay networks},'' in \emph{ACM SIGCOMM Computer
  Communication Review}, vol.~32, no.~4.\hskip 1em plus 0.5em minus 0.4em\relax
  ACM, 2002.

\bibitem{dimakis2006distributed}
A.~Dimakis, V.~Prabhakaran, and K.~Ramchandran, ``{Distributed fountain codes
  for networked storage},'' in \emph{IEEE ICASP}, 2006.

\bibitem{cataldi2009corp}
P.~Cataldi, A.~Tomatis, G.~Grilli, and M.~Gerla, ``{CORP: Cooperative rateless
  code protocol for vehicular content dissemination},'' in \emph{IFIP
  Med-Hoc-Net}, 2009.

\bibitem{bogino2007sliding}
M.~Bogino, P.~Cataldi, M.~Grangetto, E.~Magli, and G.~Olmo, ``{Sliding-window
  digital fountain codes for streaming of multimedia contents},'' in \emph{IEEE
  ISCAS}, 2007.

\bibitem{cataldi2010sliding}
P.~Cataldi, M.~Grangetto, T.~Tillo, E.~Magli, and G.~Olmo, ``{Sliding-window
  raptor codes for efficient scalable wireless video broadcasting with unequal
  loss protection},'' \emph{IEEE TIP}, vol.~19, no.~6, 2010.

\bibitem{bttf}
M.~Dell'Amico, P.~Michiardi, and Y.~Roudier, ``Back to the future: On
  predicting user uptime,'' \emph{CoRR}, vol. abs/1010.0626, 2010.

\bibitem{mager09}
T.~Mager, E.~Biersack, and P.~Michiardi, ``A measurement study of the wuala
  on-line storage service,'' in \emph{Peer-to-Peer Computing (P2P), 2012 IEEE
  12th International Conference on}, sept. 2012, pp. 237 --248.

\bibitem{p2p11}
L.~{T}oka, M.~{D}ell'Amico, and P.~{M}ichiardi, ``Data transfer scheduling for
  p2p storage,'' in \emph{IEEE P2P}, 2011.

\bibitem{Piatek07doincentives}
M.~Piatek, T.~Isdal, T.~Anderson, A.~Krishnamurthy, and A.~Venkataramani, ``Do
  incentives build robustness in {BitTorrent}?'' in \emph{USENIX NSDI}, 2007.

\bibitem{batten2001pstore}
C.~Batten, K.~Barr, A.~Saraf, and S.~Trepetin, ``{pStore: A secure peer-to-peer
  backup system},'' \emph{Unpublished report, MIT Laboratory for Computer
  Science}, 2001.

\bibitem{cox02pastiche}
L.~Cox and B.~Noble, ``Pastiche: Making backup cheap and easy,'' in
  \emph{USENIX OSDI}, 2002.

\bibitem{1021936}
M.~Landers, H.~Zhang, and K.-L. Tan, ``Peerstore: Better performance by
  relaxing in peer-to-peer backup,'' in \emph{IEEE P2P}, 2004.

\bibitem{Sameh03acooperative}
M.~Lillibridge, S.~Elnikety, A.~Birrell, M.~Burrows, and M.~Isard, ``A
  cooperative internet backup scheme,'' in \emph{USENIX ATC}, 2003.

\bibitem{Haeberlen05glacier}
A.~Haeberlen, A.~Mislove, and P.~Druschel, ``Glacier: Highly durable,
  decentralized storage despite massive correlated failures,'' in \emph{USENIX
  NSDI}, 2005.

\bibitem{Chun06efficientreplica}
B.-g. Chun, F.~Dabek, A.~Haeberlen, E.~Sit, H.~Weatherspoon, M.~F. Kaashoek,
  J.~Kubiatowicz, and R.~Morris, ``Efficient replica maintenance for
  distributed storage systems,'' in \emph{USENIX NSDI}, 2006.

\bibitem{Kubiatowicz00oceanstore}
J.~Kubiatowicz, D.~Bindel, Y.~Chen, S.~Czerwinski, P.~Eaton, D.~Geels,
  R.~Gummadi, S.~Rhea, H.~Weatherspoon, W.~Weimer, C.~Wells, and B.~Zhao,
  ``Oceanstore: An architecture for global-scale persistent storage,'' in
  \emph{ACM ASPLOS}, 2000.

\bibitem{pamies2010availability}
L.~Pamies-Juarez, P.~García-López, and M.~S{\'a}nchez-Artigas,
  ``Availability and redundancy in harmony: Measuring retrieval times in p2p
  storage systems,'' in \emph{IEEE P2P}, 2010.

\bibitem{oualha2008security}
N.~Oualha, M.~\"Onen, and Y.~Roudier, ``{A security protocol for
  self-organizing data storage},'' in \emph{IFIP SEC}, 2008.

\bibitem{ateniese2008scalable}
G.~Ateniese, R.~Di~Pietro, L.~Mancini, and G.~Tsudik, ``Scalable and efficient
  provable data possession,'' in \emph{ICST SecureComm}, 2008.

\bibitem{kamvar2003eigentrust}
S.~Kamvar, M.~Schlosser, and H.~Garcia-Molina, ``The {EigenTrust} algorithm for
  reputation management in {P2P} networks,'' in \emph{ACM WWW}, 2003.

\bibitem{vishnumurthy2003karma}
V.~Vishnumurthy, S.~Chandrakumar, and E.~Sirer, ``Karma: A secure economic
  framework for peer-to-peer resource sharing,'' in \emph{P2P Econ}, 2003.

\bibitem{Pamies-juarez_rewardingstability}
L.~Pamies-Juarez, P.~Garc\'ia-L\'opez, and M.~S\'anchez-Artigas, ``Rewarding
  stability in peer-to-peer backup systems,'' in \emph{IEEE ICON}, 2008.

\bibitem{EURECOM+2738}
P.~{M}ichiardi and L.~{T}oka, ``{S}elfish neighbor selection in peer-to-peer
  backup and storage applications,'' in \emph{Euro-Par}, 2009.

\bibitem{Tangwongsan_efficientsimilarity}
K.~Tangwongsan, H.~Pucha, D.~G. Andersen, and M.~Kaminsky, ``Efficient
  similarity estimation for systems exploiting data redundancy,'' in \emph{IEEE
  INFOCOM}, 2010.

\bibitem{EURECOM+3140}
L.~{T}oka, M.~{D}ell'Amico, and P.~{M}ichiardi, ``{O}nline data backup: a
  peer-assisted approach,'' in \emph{IEEE P2P}, 2010.

\end{thebibliography}

\end{document}